\documentclass[12pt,epsf]{article}
\pdfoutput=1
\usepackage{amssymb,amsmath}
\usepackage{graphicx, xcolor, varwidth}
\usepackage[percent]{overpic}
\usepackage{setspace}

\definecolor{mlightgray}{gray}{0.8}
\colorlet{darkblue}{blue!70!black}
\usepackage[colorlinks=true, urlcolor=darkblue,linktocpage=true,linkcolor=darkblue,citecolor=darkblue]{hyperref}

\numberwithin{equation}{section}

\newcommand{\be}{\begin{equation}}
\newcommand{\ee}{\end{equation}}
\newcommand{\bea}{\begin{eqnarray}}
\newcommand{\eea}{\end{eqnarray}}
\newcommand{\bear}{\begin{eqnarray}}
\newcommand{\eear}{\end{eqnarray}}
\newcommand{\beas}{\begin{eqnarray*}}
\newcommand{\p}{\partial}
\newcommand{\eeas}{\end{eqnarray*}}
\newcommand{\ba}{\begin{array}}
\newcommand{\ea}{\end{array}}

\newcommand{\pd}[2][1]{\ifnum#1=1 \frac{\partial}{\partial {#2}} \else
  \frac{\partial^#1}{\partial {#2}^{#1}}\fi}
\newcommand{\dpd}[2][1]{\ifnum#1=1 \dfrac{\partial}{\partial {#2}} \else
  \frac{\partial^#1}{\partial {#2}^{#1}}\fi}
\newcommand{\td}[2][1]{\ifnum#1=1 \frac{d}{d{#2}} \else
  \frac{d^#1}{d{#2}^{#1}}\fi}

\newcommand{\nbox}{{\,\lower0.9pt\vbox{\hrule \hbox{\vrule height 0.2 cm \hskip 0.19 cm \vrule height 0.2 cm}\hrule}\,}}
\newcommand{\Tr}{\ {\rm Tr}\ }

\newcommand{\ie}{{\it i.e.,}\ }

\newcommand{\ct}{\frac{c}{12}}
\newcommand{\cb}{\frac{c}{24}}

\newcommand{\ZB}[1]{Z[#1]}
\newcommand{\ZBp}[1]{Z'[#1]}
\newcommand{\btau}{\bar{\tau}}

\newcommand{\ec}{u}

\newcommand{\C}{\mathcal{C}}
\newcommand{\Z}{\mathcal{Z}}
\newcommand{\bq}{\bar{q}}
\newcommand{\bh}{\bar{h}}
\renewcommand{\c}[1]{\frac{c}{#1}}

\newcommand{\cS}{\mathcal{S}}

\newcommand{\THecke}{T}
\newcommand{\dT}[1]{d_{T_{#1}}}

\begin{document}
\begin{spacing}{1.3}
\begin{titlepage}

\begin{center}
{\Large \bf Universal Spectrum of 2d Conformal Field Theory\\ 
\vspace{.4cm} in the Large $c$ Limit}

\vspace*{6mm}

Thomas Hartman,$^*$ Christoph A.~Keller,$^\dagger$ and Bogdan Stoica$^\ddag$

\vspace{5mm}

\textit{
$^*$ Kavli Institute for Theoretical Physics, University of California\\
Santa Barbara, CA 93106-4030 USA}\\ 

\vspace{2mm}

\textit{
$^\dagger$ NHETC, Rutgers, The State University of New Jersey\\
Piscataway, NJ 08854-8019 USA}\\ 

\vspace{2mm}

\textit{
$^\ddag$ Walter Burke Institute for Theoretical Physics,\\ California Institute of Technology, 452-48, Pasadena, CA 91125, USA}\\ 

\vspace{5mm}

{\tt  thartman@kitp.ucsb.edu, keller@physics.rutgers.edu,  bstoica@theory.caltech.edu }

\vspace*{1cm}
\end{center}
\begin{abstract}

Two-dimensional conformal field theories exhibit a universal free energy in the high temperature limit $T\to\infty$, and a universal spectrum in the Cardy regime, $\Delta \to \infty$. We show that a much stronger form of universality holds in theories with a large central charge $c$ and a sparse light spectrum. In these theories, the free energy is universal at all values of the temperature, and the microscopic spectrum matches the Cardy entropy for all $\Delta \geq \frac{c}{6}$.  The same is true of three-dimensional quantum gravity; therefore our results provide simple necessary and sufficient criteria for 2d CFTs to behave holographically in terms of the leading spectrum and thermodynamics. We also discuss several applications to CFT and gravity, including operator dimension bounds derived from the modular bootstrap, universality in symmetric orbifolds, and the role of non-universal `enigma' saddlepoints in the thermodynamics of 3d gravity.

\end{abstract}
\vspace{1cm}

\begin{flushleft}{\small CALT 68-2889, RUNHETC-2014-07}\end{flushleft}

\end{titlepage}
\end{spacing}

\vskip 1cm

\tableofcontents

\begin{spacing}{1.3}

\section{Introduction}

In quantum gravity different energy scales do not decouple in the same way as in standard effective field theory. Rather, as a consequence of diffeomorphism invariance, the theory in the UV is heavily constrained by the IR. The same effect must occur in conformal field theories (CFTs) with holographic duals.  In this paper we explore this connection in a class of 2d CFTs, where it is realized as invariance under large conformal transformations of the theory on a torus, and provide a partial answer to the question of what data in the UV is fixed by the IR.  The results agree with known universal features of 3d gravity. The calculations are entirely within CFT  and do not assume holography.

The UV/IR connection leads to universality.  A famous example in gravity is black hole entropy: to leading order, every UV theory governed by the Einstein action at low energies has the same high energy density of states, dictated by the Bekenstein-Hawking entropy law $S = \mbox{Area}/4G_{N}$. This is an IR constraint on the UV completion. The area law has been derived in great detail for particular black holes in string theory \cite{Strominger:1996sh}. Yet it is often mysterious in these calculations why the final answer is simple and universal, since the intermediate steps seem to rely on various UV details.

In $AdS_3$ gravity, the black hole entropy agrees with the Cardy formula \cite{Cardy:1986ie} for the asymptotic density of states in any unitary, modular invariant 2d CFT \cite{Strominger:1997eq}:
\be\label{introcardy}
S_{black\ hole}(E_L, E_R) = S_{Cardy}(E_L, E_R) \equiv 2 \pi \sqrt{\c{6}E_L} + 2 \pi \sqrt{\c{6}E_R} \ .
\ee
The central charge takes the Brown-Henneaux value \cite{Brown:1986nw}, 
\be
c = \frac{3\ell}{2G_N} \gg 1 \ ,
\ee
where $\ell$ is the $AdS$ radius, $G_N$ is Newton's constant, and $E_{L,R}$ are the left- and right-moving energies of the black hole (normalized so that the vacuum has $E_L = E_R = -\c{24}$). This is a more universal derivation of the black hole entropy that does not rely on all of the microscopic details of the CFT.  However, there is an important difference between the black hole entropy and the Cardy formula. In general the Cardy formula only holds in the Cardy limit
\be
c \quad \textrm{fixed}\ ,\qquad E_{L,R} \rightarrow \infty\ ,
\ee
whereas the Bekenstein-Hawking entropy should hold in a semiclassical limit,
\be\label{BHregime}
c \rightarrow \infty, \qquad E_{L,R} \sim c \ .
\ee
Having an extended range of validity of the Cardy formula is a key feature that distinguishes holographic CFTs from the rest.  Of course, in the explicit theories considered in \cite{Strominger:1996sh,Strominger:1997eq}, it is possible to check microscopically that the Cardy formula indeed applies beyond its usual range, but in other cases such as the Kerr/CFT correspondence the Cardy formula is applied without a clear justification \cite{Guica:2008mu}.

One aim of the present paper is to characterize the class of CFTs in which the Cardy formula (\ref{introcardy}) extends to the regime (\ref{BHregime}). It is often stated that this should be the case in a theory with a `large gap' in operator dimensions above zero.\footnote{Not to be confused with another common statement that it may apply when there is a `small gap' above the black hole threshold (discussed for example in \cite{Guica:2008mu}) suggesting a long string picture. We will not address this latter criterion.}  We confirm this intuition, give precise necessary and sufficient criteria, and identify the applicable range of $E_{L,R}$. The origin of the UV/IR connection in 2d CFT is modular invariance, so this is our starting point. In terms of the partition function at inverse temperature $\beta$, the modular $S$-transformation implies
\be\label{modinv}
Z(\beta) = Z(\frac{4\pi^2}{\beta}) \ .
\ee
 The standard Cardy formula was derived by taking $\beta \to 0$ in this formula, so it is valid in the small-$\beta$ limit at any value of $c$ \cite{Cardy:1986ie}.  We will essentially repeat the analysis in the limit $c \to \infty$ with $\beta$ held fixed. The result is the same formula for $Z(\beta)$, but valid in the large $c$ limit at any value of $\beta$, under certain conditions on the light spectrum in addition to the usual assumptions of unitarity and modular invariance. This is the limit that applies to 3d black holes.

Constraints from modular invariance have been studied extensively in the simplified settings of holomorphic CFT and rational CFT.  In the holomorphic case, with only left-movers, the partition function $Z(\tau)$ is a holomorphic function of the complexified temperature $\tau$. For a given central charge, the space of holomorphic partition functions is finite dimensional, which yields powerful constraints. For example, the spectrum of states with $E_L > 0$ is uniquely fixed by the spectrum with $E_L \leq 0$, and there must be at least one primary operator in the range $-\c{24} < E_L \leq \c{24} + 1$. Similar statements apply to other holomorphic objects such as BPS partition functions and elliptic genera in supersymmetric theories (see for example \cite{Dijkgraaf:2000fq,Manschot:2007ha,Witten:2007kt,Gaberdiel:2008xb}). Far less is known about modular invariance in non-holomorphic theories. For some rational CFTs, the solutions of (\ref{modinv}) can be classified explicitly \cite{Cappelli:1986hf}. For general non-rational partition functions, one of the only tools beyond the Cardy formula is the modular bootstrap \cite{Hellerman:2009bu}, in which (\ref{modinv}) is expanded order by order around the self-dual temperature $\beta=2\pi$. We use our methods to reproduce and clarify some results of the bootstrap in section \ref{ss:opbounds}. This indicates that a large $c$ expansion may be a useful way to organize the constraints of modular invariance on non-holomorphic partition functions. 

This is similar in spirit to recent efforts to derive universal features of entanglement entropy \cite{Hartman:2013mia,Barrella:2013wja,Chen:2013kpa,Perlmutter:2013paa} and gravitational interactions \cite{Fitzpatrick:2014vua} at large $c$.  In fact, since the second Renyi entropy of two disjoint intervals can be conformally mapped to the torus partition function at zero angular potential, the entanglement entropy is directly related.  Most of the entanglement calculations rely on a small interval expansion, but our results do not, so this rules out the possibility of missing saddlepoints in the second Renyi entropy discussed in \cite{Hartman:2013mia,Faulkner:2013yia}. Under what conditions universality holds for higher genus partition functions (or higher Renyi entropies) is an important open question.

\subsection{Summary of results}

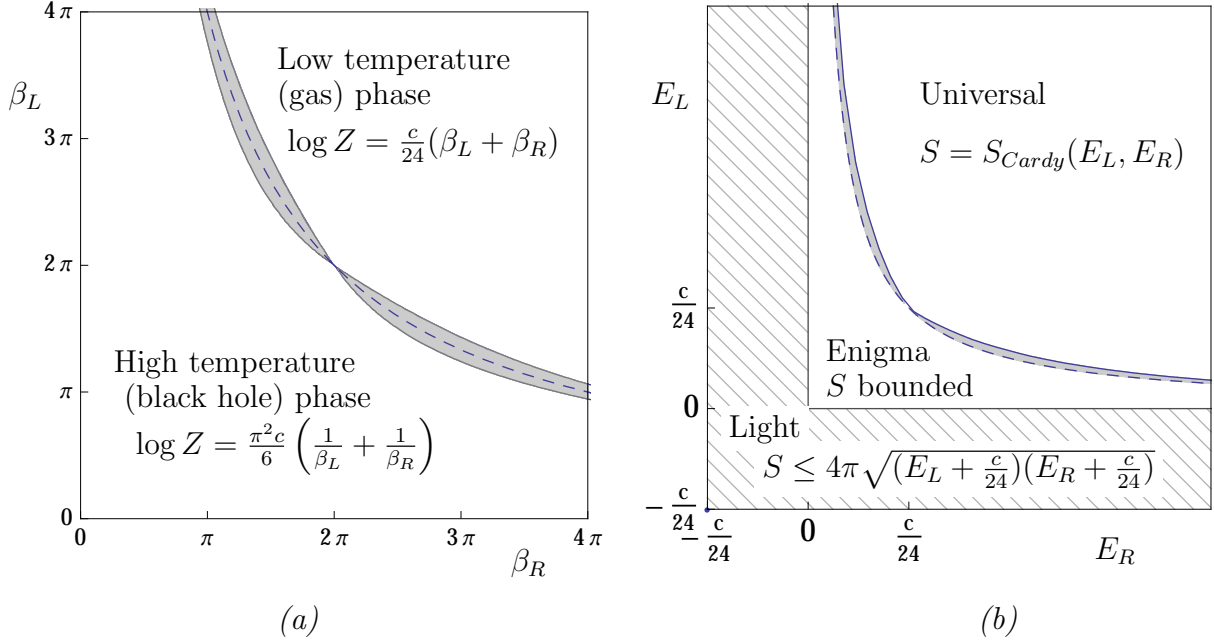
\begin{figure}[t]
\center

\begin{overpic}{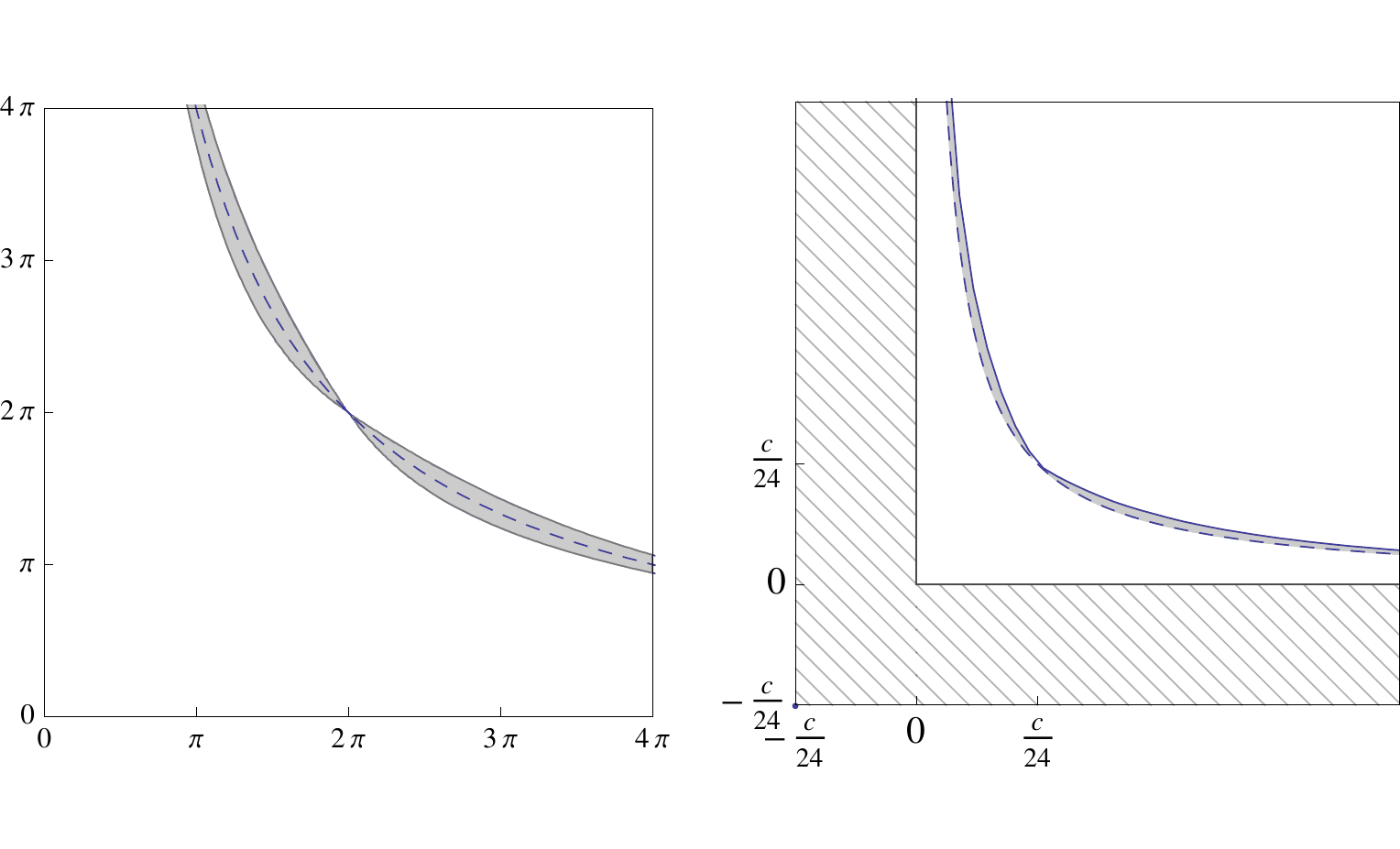}

\put(20, 0) {\textit{(a)}}

\put (-3,45) {$\beta_L$}
\put (40,5) {$\beta_R$}
\put (20,48) {Low temperature}
\put (20,45) {(gas) phase}
\put (20,41) { $\log Z = \c{24}(\beta_L+\beta_R)$}
\put (6,22) {High temperature}
\put(6,19) { (black hole) phase}
\put (7,15) { $\log Z = \frac{\pi^2 c}{6}\left( \frac{1}{\beta_L} + \frac{1}{\beta_R}\right)$}

\put(80,0) {\textit{(b)}}

\put(90,6) {$E_R$}
\put(52,45) {$E_L$}
\put (75, 45) {Universal}
\put (75, 40) {$S = S_{Cardy}(E_L,E_R)$}
\put (60,13) { \colorbox{white}{$S \leq 4\pi \sqrt{(E_L + \c{24})(E_R + \c{24})}$}}
\put (58,16.5) {\colorbox{white}{Light}}
\put (67,23) {Enigma}
\put (67,20) {$S$ bounded}

\end{overpic}
\caption{Universality in CFT with large $c$ and a sparse light spectrum. \textit{(a) Canonical Ensemble: } The dashed line ($\beta_L\beta_R = 4\pi^2$) separates high temperatures from low temperatures; in gravity, this would be the Hawking-Page phase transition. We show that the leading free energy is universal and equal to the Cardy value outside of the shaded sliver, and conjecture that this also holds in the sliver.    \textit{(b) Spectrum:} The density of light states in the hatched region is bounded above by the sparseness assumption.  We show that the density of states obeys the Cardy formula above the solid curve, and conjecture that this is true above the dashed curve ($E_LE_R = (c/24)^2$). In the enigma range, the entropy is not universal, but satisfies an upper bound that prevents the enigma states from dominating the canonical ensemble.
 \label{fig:phases}}
\end{figure}

Operators in a unitary 2d CFT are labeled by their left and right conformal weights
$(h, \bh)$ with $h,\bh \geq 0$ . If we put the theory
on a circle of length $2\pi$, the operator-state correspondence
associates to each operator a state with energies
\be
E_L = h - \c{24} \ , \quad\quad E_R = \bh - \c{24} \  
\ee
and total energy 
\be
E = E_L + E_R = \Delta - \c{12} \ .
\ee
In section \ref{s:canonical} we study the partition function for zero angular potential,
\be
Z(\beta) = \sum e^{-\beta E}\ .
\ee
It is convenient to classify states as \textit{light}, \textit{medium}, or \textit{heavy}:
\be
\mbox{light}: -\c{12}\leq E \leq \epsilon\ , \qquad
\mbox{medium}: \epsilon < E < \c{12} \ , \qquad
\mbox{heavy}: E \geq \c{12} \ ,
\ee
for some small positive number $\epsilon$ that is eventually taken to zero in the large $c$ limit.  We show that the free energy is fixed up to small corrections by the light spectrum.  
If in addition we also assume that the spectrum of light states is sparse, 
by which we mean that it is bounded as
\be\label{introasum}
\rho(E) = \exp[S(E)] \lesssim \exp\left[ 2\pi \left(E+\frac{c}{12}\right)\right] \ , \qquad E \leq \epsilon
\ee
then at large $c$ the free energy is universal to leading order :
\be
\log Z(\beta)= \c{12} \max\left(\beta, \frac{4\pi^2}{\beta}\right) + O(c^0)\ .
\ee
There is a phase transition at $\beta=2\pi$. Furthermore the microscopic spectrum satisfies the Cardy formula for all heavy states,
\be
S(E) \sim 2\pi \sqrt{\c{3}E} \qquad (E \geq \c{12}) \ .
\ee
The medium-energy regime does not have a universal entropy, but it is bounded by
\be\label{introbbound}
S(E) \lesssim \frac{\pi c}{6} + 2 \pi E \qquad (\epsilon < E < \c{12}) \ .
\ee
The medium-energy states never dominate the canonical ensemble and therefore do not affect the leading free energy.

The heavy states are holographically dual to stable black holes.
The non-universal entropy at medium energies is related to the fact that in 3d gravity, black holes in this range are thermodynamically unstable.  In fact, the leading order spectrum of 3d gravity plus matter (or gravity on $AdS_3 \times X$) in this range is also non-universal, because in addition to the usual BTZ black holes there can be entropically dominant `enigmatic' black holes \cite{deBoer:2008fk,Bena:2011zw}.  These solutions, discussed in section \ref{s:gravity}, obey the bound (\ref{introbbound}).

In section \ref{s:angularpot} we repeat the analysis for non-zero angular potential, which
means we introduce $\beta_L$ and $\beta_R$.  
The partition function at finite temperature and angular potential is
\be\label{fullz}
Z(\beta_L, \beta_R) = \sum e^{-\beta_L E_L - \beta_R E_R} \ .
\ee
The results are more intricate but qualitatively similar, and summarized in figure \ref{fig:phases}. 
In the quadrants $\beta_{L}, \beta_R>2\pi$ and $\beta_L,\beta_R<2\pi$, the free energy is universal assuming a sparse light spectrum (\ref{introasum}). If we further restrict the mixed density of states as
\be\label{introamix}
\rho(E_L,E_R) \lesssim \exp\left[4\pi \sqrt{(E_L+\c{24})(E_R+\c{24})} \right]  \qquad (E_L<0 \mbox{\ or\ } E_R<0) \ ,
\ee
then we can show that the universal behavior 
\be\label{introFangular}
\log Z(\beta_L,\beta_R)= \c{24} \max\left(\beta_L+\beta_R, \frac{4\pi^2}{\beta_L}+\frac{4\pi^2}{\beta_R}\right) + O(c^0)
\ee
extends to the rest of the $(\beta_L,\beta_R)$ plane outside
of a small sliver near the line $\beta_L\beta_R = 4\pi^2$.
The universal features of the free energy lead to corresponding universal features of the entropy $S(E_L,E_R)$; it equals $S_{Cardy}(E_L,E_R)$ at high enough energies, and is bounded above in the intermediate range (see figure \ref{fig:phases}b).
 The derivation of the free energy is an iterative procedure that gradually eliminates larger portions of the $(\beta_L,\beta_R)$ plane. The sliver shown in the figure is what remains after three iterations, but we conjecture that more iterations would show that the free energy is universal for all  $\beta_L\beta_R \neq 4\pi^2$.  If so, then the Cardy entropy formula holds for all $E_L E_R > \left(\c{24}\right)^2$.

The detailed comparison to 3d gravity is made in section \ref{s:gravity}.  Finally in section \ref{s:orbifold} we compare our results to symmetric orbifold CFTs, since certain symmetric orbifolds are known to have holographic duals.  We show that all symmetric orbifolds have free energy that satisfies (\ref{introFangular}) at all temperatures. We also show that the leading behavior of the density of states is completely universal for all symmetric orbifold theories, and saturates the bounds (\ref{introasum}), (\ref{introbbound}) and (\ref{introamix}).  In this sense, symmetric orbifolds have the maximally dense spectrum compatible with 3d gravity.

\section{The large $c$ partition function}\label{s:canonical}
\subsection{Setup}
We begin by analyzing the constraints of modular invariance on the partition function at zero angular potential, $\beta_L = \beta_R = \beta$.  Modular invariance requires
\be
Z(\beta) = Z(\beta') \ , \qquad \beta' \equiv \frac{4\pi^2}{\beta} \ .
\ee
We denote the light states by $L$, and the medium and heavy states by $H$,
\be
L =\{ E \leq \epsilon \}  \  , \qquad H =\{ E > \epsilon \}\ ,
\ee
and define the corresponding contributions to the partition
function and its dual 
in the obvious way,
\begin{align}
\ZB{L} &= \sum_L e^{-\beta E}  &  \ZB{H} &= \sum_H e^{-\beta E} \label{defzbrak}\\
\ZBp{L} &= \sum_L e^{-\beta' E} &  \ZBp{H} &= \sum_H e^{-\beta' E} \ .\notag
\end{align}
Clearly the full partition function is
\be
Z(\beta) = \ZB{L} + \ZB{H} = \ZBp{L} + \ZBp{H} \ .
\ee

\subsection{Free energy}
Let us first discuss to what extent the
light spectrum determines the free energy. As pointed
out in the introduction, in
the holomorphic case, it is completely determined by $L$. In the non-holomorphic case,
clearly for very small temperature it is given
by the light states, or more precisely, by the vacuum.
For very high temperature 
we know from the usual Cardy formula that
the behavior is again determined by the vacuum
via modular invariance. We want to investigate
what we can say about intermediate temperatures
assuming that we know $L$ completely.

We can express modular invariance as
\be
\ZB{L}-\ZBp{L}=\ZBp{H}-\ZB{H}\ .
\ee
In a first step we want to bound $\ZB{H}$.
Assume $\beta > 2\pi$. Then
\be
\ZB{H} = \sum_{E>\epsilon} e^{(\beta'-\beta)E}e^{-\beta'E} \leq e^{(\beta'-\beta)\epsilon}\, \ZBp{H}\ .
\ee
Therefore we have
\be
-\ZBp{H}(1-e^{(\beta'-\beta)\epsilon})\geq \ZB{H}-\ZBp{H} \ .
\ee
Using modular invariance,
\begin{align}
\ZBp{H} \leq (1-e^{(\beta'-\beta)\epsilon})^{-1}&(\ZBp{H} -\ZB{H})\\
&=(1-e^{(\beta'-\beta)\epsilon})^{-1}(\ZB{L}-\ZBp{L})
\leq (1-e^{(\beta'-\beta)\epsilon})^{-1}\ZB{L}\ ,\notag
\end{align}
so in total we have
\be \label{Hbound}
\ZB{H} \leq \frac{e^{(\beta'-\beta)\epsilon}}{1-e^{(\beta'-\beta)\epsilon}}\ZB{L}\ .
\ee
So for $\beta>2\pi$ we have for the free energy
\be\label{lightsandwich}
\log \ZB{L} \leq \log Z \leq 
 \log \ZB{L} - \log (1-e^{(\beta'-\beta)\epsilon})\ .
\ee
By modular invariance we obtain an analogous expression
for $\beta < 2\pi$.

The two inequalities in (\ref{lightsandwich}) tell us
that the free energy of a theory differs from the
contribution of the light states only within a universal range which does not depend on the theory.
Crucially however this error is not bounded
uniformly in $\beta$. The closer the temperature
is to the self-dual point (and the smaller we
choose $\epsilon$ for that matter), the bigger
an error we make. For $\beta=\beta'$ in particular
we can only give a lower bound for the free energy.

Let us now consider families of CFTs depending on the
central charge $c$, and investigate the limit of large $c$.
From (\ref{lightsandwich}) we can obtain the free
energy of this family as
\be\label{zuniv}
\log Z(\beta) = \left\{ \begin{array}{ccc} \log \ZB{L} + O(1) &: &\beta > 2\pi\\
\log \ZBp{L} +O(1) &:&  \beta < 2\pi \end{array} \right. \ 
\ee
in the limit $c\rightarrow\infty$. We stress again
that the error is not uniform in 
$\beta$: for large but finite $c$, we can always
find $\beta$ close enough to $2\pi$ so that the
$O(1)$ term is potentially of the same order
as the light state contribution.

This result is particularly powerful in a theory where the $\ZB{L}$ is dominated by the vacuum state.  In this case
\be\label{vaconly}
\log Z(\beta) = \left\{ \begin{array}{ccc} \frac{c}{12}\beta   + O(1)&:& \beta > 2\pi\\
\frac{\pi^2 c}{3\beta} + O(1)&:&  \beta < 2\pi \end{array} \right. \ .
\ee
It is straightforward to see that this holds if and only if
\be\label{lightgrowth}
\log \left(1+\sum_{0<\Delta\leq c/12+\epsilon} e^{-\beta\Delta}\right) = O(1) \ ,
\ee
for $\beta > 2\pi$. Allowing for $o(c)$ corrections to the free energy, we can also choose to take $\epsilon \to 0$ in the large $c$ limit (for example $\epsilon \sim e^{-\alpha \sqrt{c}}$ for some $\alpha>0$), and the conclusion is that the free energy is universal if and only if the density of light states satisfies\footnote{Approximation symbols are used with precise definitions: $x \sim y$ means $\lim x/y = 1$, $x \approx y$ means $\lim \frac{\log x}{\log y} = 1$, and depending on the context, inequalities $x \lesssim y$ mean $\lim x/y \leq 1$ if $x = O(c)$ (for example a free energy) or $\lim \frac{\log x}{\log y} \leq 1$ for exponential quantities (partition functions).}
\be\label{canbound}
\rho(E) \lesssim \exp\left[2\pi (E+\c{12})\right]  \qquad (E \leq  \epsilon)\ .
\ee

\subsection{Spectrum}
Let us now discuss what we can learn about the heavy spectrum of the theory
from (\ref{vaconly}). Thermodynamically this means we are
interested in the entropy $S(E)$. This we can obtain by performing
the standard Legendre transform from $F(\beta)$ to $E(S)$.
By the usual arguments, $F(\beta)$ fixes $E(S)$ completely,
so naively we could expect that (\ref{vaconly})
gives the leading $c$ behavior of $S(E)$. It turns out
that is not the case, and that subleading corrections
to $F$ can give large $c$ corrections to $S(E)$,
so that we can only fix the leading order behavior
of $S(E)$ in a certain range of $E$.

To see this more concretely, we compute the thermodynamic energy
\be
E(\beta) = -\p_\beta \log Z = \left\{ \begin{array}{ccc} -\frac{c}{12}+O(1) &:& \beta>2\pi\\
\frac{\pi^2 c}{3\beta^2}+O(1) &:& \beta < 2\pi \end{array}\right. \ .
\ee
and thermodynamic entropy
\be
S(\beta) = (1-\beta \p_\beta)\log Z =  \left\{ \begin{array}{ccc} O(1) &:& \beta>2\pi\\
\frac{2\pi^2 c}{3\beta}+O(1) &:& \beta < 2\pi \end{array}\right. \ .
\ee
We see that at $\beta = 2\pi$, $E$ jumps from $-\ct$ to $\ct$.
For finite $c$ of course $E$ has to be regular. What
this means is that a small change of order $O(1)$ 
in $\log Z$ at $\beta\sim 2\pi$ will produce a change of order $c$ in $E$.
This is the flip side of (\ref{lightsandwich}) which
tells us that we should only trust our approximations
if $\beta$ is far enough from the self-dual temperature.
For the microcanonical density of states, this means that we should
only trust our approximation if $E$ is in the stable region $>\ct$.
In that case we get the expected Cardy behavior
\be\label{sereg}
S(E) \sim 2\pi \sqrt{\frac{c}{3}E} \quad\quad (E > \frac{c}{12}) \ .
\ee
This entropy was obtained from thermodynamics, but it also holds for the microscopic density of states,
\be
\rho(E) \approx e^{S(E)} \ .
\ee 
This is expected since $c \to \infty$ behaves like a thermodynamic limit, but as usual it requires some averaging to make precise.  The details are relegated to appendix \ref{app:micro}.

\subsection{Subleading saddles and the enigmatic range}\label{ss:cftsubsaddles}
For reasons that will be clear when we compare to 3d gravity, we refer to the medium-energy states
\be
0 < E < \c{12}
\ee
as the `enigmatic' range. The saddlepoint that dominates the partition function at large $c$ never falls in this range, so $S(E)$ is not universal.  We can, however, easily derive an upper bound.  Setting $\beta = 2\pi$ in the expression $Z(\beta) > \rho(E) e^{-\beta E}$ gives
\be\label{eupi}
S(E) \lesssim \frac{\pi c}{6} + 2\pi E \ .
\ee
This holds universally in theories obeying (\ref{lightgrowth}). We have not found a universal lower bound --- in particular, our results and the results in \cite{Hellerman:2009bu,Friedan:2013cba} seem to be compatible with the possibility that there are no primary states within this range --- but modular invariance suggests a lower bound may hold in many theories.  To see this, write the contribution of heavy states to the partition function as
\be\label{reorg}
\ZB{H} = \ZBp{L} + \big( \ZBp{H} - \ZB{L} \big) \ .
\ee
For $\beta > 2\pi$, the terms in parentheses dominate.  Still, there is a contribution to the first term from the vacuum state, 
\be
\ZB{H}  = e^{\c{12}\beta'} + \cdots \ .
\ee
If the heavy spectrum is precisely tuned so the dominant terms in parentheses cancel this contribution, then $\ZB{H}$ is completely unknown.  If on the other hand we assume this cancellation does not happen then we expect a corresponding contribution to the density of states, $S(E) \sim 2\pi\sqrt{\c{3}E} + \cdots$.  This suggests that in generic theories without fine tuning the entropy in the enigmatic range also satisfies a lower bound,
\be\label{specup}
2\pi \sqrt{\c{3}E} \ \  \lesssim \ \  S(E) \quad \lesssim\quad  \frac{\pi c}{6} + 2\pi E \qquad (0<E<\c{12}) \ .
\ee
As we will see in section~\ref{s:orbifold}, there are theories which saturate the upper bound of (\ref{specup}).
We can also construct leading order partition functions which saturate the lower bound: Take for instance the partition function
whose light spectrum only contains the vacuum representation, and whose heavy state contribution is given by $\ZB{H}:=\ZBp{L} + $ subleading.
We do not know of any examples which have fewer medium states than this. 
This certainly does not constitute a proof, and it may be possible to evade the lower bound if the heavy spectrum can be arranged to produce delicate cancellations with the light spectrum.

\subsection{Operator bounds}\label{ss:opbounds}

As mentioned in the introduction, the light spectrum of general CFTs can also be constrained by the modular bootstrap. The idea of the modular bootstrap is to expand the partition function around the self-dual temperature $\beta = 2\pi$ and then check (\ref{modinv}) order by order.  In \cite{Hellerman:2009bu}, this technique was used to lowest order to prove that every CFT has a state with scaling dimension $\Delta_1 = E_L + E_R +\c{12} \leq \c{6} + 0.474\dots$. Other arguments such as extrapolating the result for holomorphic CFTs suggest that a tighter bound $\Delta_1 \sim \c{12}$ may be possible. A more systematic numerical analysis of the modular bootstrap at relatively large values of $c$ in \cite{Friedan:2013cba} reproduces however the same asymptotic result, 
\be\label{statemop}
\Delta_1 \lesssim \c{6} \ .
\ee
In our approach, this bound follows immediately from the fact that (\ref{sereg}) is reliable microscopically. Here the reason that the bound is $\c{6}$ and not $\c{12}$ is that the states with $\c{12} < \Delta < \c{6}$ never dominate the canonical ensemble. Our uncertainty about the medium-energy states (\ref{eupi}) thus translates exactly into an uncertainty about the best possible bound.

States above the lightest primary were incorporated into the modular bootstrap in \cite{Qualls:2013eha}. Based on the pattern observed numerically, it was conjectured that there are actually an exponentially large number of primaries at or below $\Delta \sim \c{6}$ as $c\to \infty$, specifically \cite{Qualls:2013eha}
\be\label{statemc}
\log N_{primaries}(\Delta \lesssim \c{6})\  \gtrsim \ \frac{\pi c}{6} \ .
\ee
For theories with a sparse light spectrum, the stronger bound 
\be\label{nprc}
\log N_{primaries}^{Cardy}(\Delta \lesssim \c{6}) \  \sim \ \frac{\pi c}{3} \ 
\ee
follows from our results, since in this case the Cardy regime extends to $\Delta\sim\c{6}$.
However, by adding a large number of light states to a sparse light spectrum we can push up the Cardy regime.  
Adding for example $\frac{\pi c}{6}(1+\alpha)$ light states at just below $E=0$ with $\alpha>0$,
the free energy is universal only for $\beta<2\pi(1-\alpha)$. It then follows that (\ref{sereg}) is valid only for $E > \c{12}(1-\alpha)^{-2}$, so that it falls beyond the range of (\ref{statemc}).

Let us therefore drop our assumption on the light spectrum and see how this relaxes the bound (\ref{nprc}). We showed that
\be
\ZB{H} \approx \ZBp{L} \qquad (\beta < 2\pi) \ .
\ee
From this we would like to extract information about the microscopic density of states at $E \lesssim \c{12}$. The associated energy is
\be\label{defe}
E(\beta) \equiv - \p_\beta \log \ZB{H} \approx \frac{4\pi^2}{\beta^2}\p_{\beta'}\log \ZBp{L} \ .
\ee
Since $\ZBp{L}$ has contributions only from $-\c{12}\leq E \lesssim 0$, 
\be
\p_{\beta'} \log \ZBp{L} \in [0, \ \c{12}] \ .
\ee
It follows from (\ref{defe}) that as $\beta \to 2\pi$, the energy $E(\beta)$ must fall in the range $[0, \c{12}]$ up to subleading corrections.  Since $\ZB{H}$ only has contributions from $E>0$, it follows that
the dominating contribution $E_0$ must satisfy 
\be
0 \lesssim E_0 \lesssim \c{12} , \qquad S(E_0) - 2\pi E_0 \sim \log \ZBp{L} \gtrsim \frac{\pi c}{6} \ ,
\ee
where the lower bound in the last inequality is the contribution of the vacuum.
The lowest $S(E_0)$ is achieved by assuming the dominant contribution comes from around $E_0 \sim 0$, so
\be
S(E_0) \gtrsim \frac{\pi c}{6} \ .
\ee
The distinction between counting states and counting primaries does not matter to leading order in $c$, so this is a derivation of (\ref{statemc}).

\section{Angular potential}\label{s:angularpot}

Let us introduce the partition function with different left- and right-moving temperatures,
\be\label{zdiff}
Z(\beta_L, \beta_R) = \Tr \, e^{-\beta_L E_L - \beta_R E_R} \ .
\ee
We take $\beta_L$ and $\beta_R$ to be real, which corresponds to a real angular potential proportional to $\beta_L-\beta_R$, and assume that the partition function is invariant under real modular transformations,
\be\label{realmod}
Z(\beta_L, \beta_R) = Z(\beta_L', \beta_R')  \  , \quad\quad \beta_L' = \frac{4\pi^2}{\beta_L} , \quad \beta_R' = \frac{4\pi^2}{\beta_R} \ .
\ee
This transformation at real temperatures is a consequence of modular invariance on the Euclidean torus.\footnote{In Euclidean signature, the angular potential is imaginary, and $Z(\tau,\btau) = Z(-1/\tau,-1/\btau)$ with $\tau = \frac{i \beta_L}{2\pi}$ complex and $\btau = \tau^*$. We may view $Z(\tau,\btau)$ as a holomorphic function on a domain in $\mathbf{C}^2$, with $\tau$ and $\btau$ independent complex numbers.  The function $f(\tau, \btau) = Z(\tau,\btau) - Z(-1/\tau,-1/\btau)$ is also holomorphic, and vanishes for $\btau = \tau^*$.  
The Weierstrass preparation theorem implies that the vanishing locus of a holomorphic function must be specified (at least locally) by a holomorphic equation $W(\tau,\bar{\tau}) = 0$.  Since $\bar{\tau} - \tau^* = 0$ is not holomorphic, it follows that $f=0$.
} 
Since we will rely on positivity, it is not straightforward to apply our argument directly to complex angular potential or to a chemical potential.

\begin{figure}[tbh]
\center

\begin{overpic}{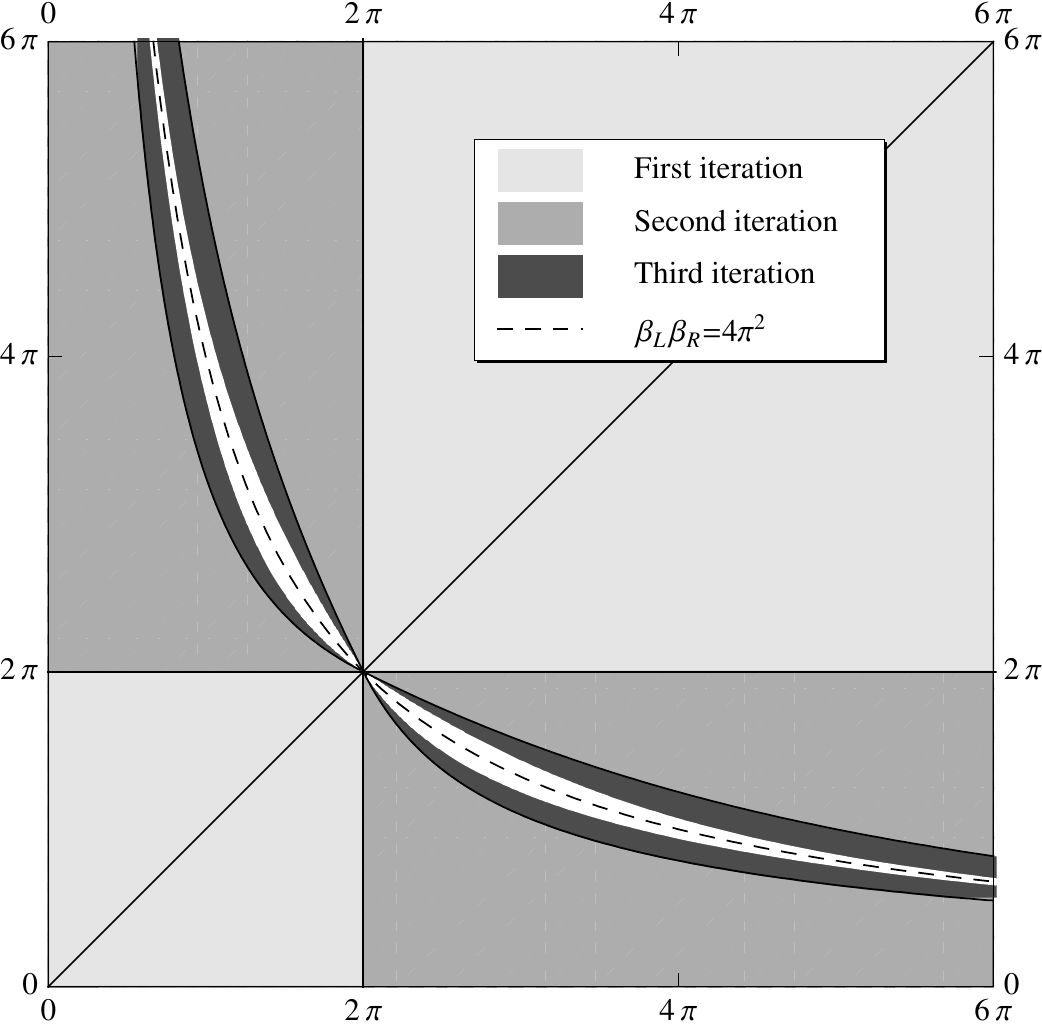}
\put (-7,80) {$\beta_L$}
\put (84,-2) {$\beta_R$}

\end{overpic}
\caption{\textit{Derivation of universal free energy at finite angular potential.} We apply an iterative procedure to derive the universal free energy in larger and larger portions of the phase diagram.  The shaded regions show the universal regions derived from the first three iterations.  After three iterations the universal range encompasses all $(\beta_L,\beta_R)$ away from the white sliver.  \label{fig:slivers}}
\end{figure}

The strategy to derive a universal free energy involves an iterative procedure, with results summarized in figure \ref{fig:slivers}.  First, we use the results of section \ref{s:canonical} to compute the free energy in the quadrants $\beta_{L,R} > 2\pi$ and $\beta_{L,R} <2\pi$.  This is then translated into new constraints on the microsopic spectrum, and used to extend the universal free energy to a larger range of $(\beta_L, \beta_R)$.  This is iterated three times. The unknown range (the white sliver in figure \ref{fig:slivers}) appears to shrink further with more iterations, so we conjecture that the universal behavior actually extends to the full phase diagram away from $\beta_L \beta_R = 4\pi^2$.

\subsection{High and low temperature partition function}
We will first discuss the regime where both temperatures
$\beta_L,\beta_R$ are either high or low. This is the region labeled `first iteration' in figure \ref{fig:slivers}. It turns out that
the constraints on the light states 
imposed in section~\ref{s:canonical} are enough to ensure
universal behavior in this regime. 
From eqs (\ref{eupi}) we know that the large $c$ density of states 
of such a theory is bounded by
\be
\rho(E_L,E_R) \leq \rho(E_L+E_R) \lesssim \exp\left(\frac{\pi c}{6} + 2\pi(E_L+E_R)\right) \ .
\ee
Therefore for $\beta_{L,R}>2\pi$, the total exponent in the partition function
\be
 \sum_{E_L,E_R}  \rho(E_L,E_R) e^{-\beta_L E_L - \beta_R E_R} 
\ee
is bounded above by
\be
\frac{\pi c}{6} + 2\pi(E_L + E_R) - \beta_LE_L - \beta_R E_R \lesssim \c{24}(\beta_L+\beta_R) \ .
\ee
This implies that the vacuum exponentially dominates over other contributions to (\ref{zdiff}) at low temperatures,
\be\label{zdifflow}
Z(\beta_L,\beta_R) \approx \exp\left[ \c{24}(\beta_L+\beta_R) \right] \quad\quad (\beta_{L,R}>2\pi) \ .
\ee
By modular invariance, we then immediately obtain at high temperatures
\be\label{zdiffhi}
Z(\beta_L,\beta_R) \approx \exp\left[ \frac{\pi^2 c}{6}\left(\frac{1}{\beta_L}+\frac{1}{\beta_R}\right)\right] \quad\quad(\beta_{L,R}<2\pi) \ .
\ee

\subsection{Spectrum}\label{ss:mixspectrum}
Just as in section~\ref{s:canonical},
the free energies (\ref{zdifflow}) and (\ref{zdiffhi}) lead to corresponding statements about the microscopic spectrum. The thermodynamic energies derived from this partition function are
\be
E_{L,R} = -\p_{\beta_{L,R}} \log Z \sim \left\{ \begin{array}{cc}
\frac{\pi^2 c}{6 \beta_{L,R}^2} & \beta_{L,R}<2\pi\\
0 & \beta_{L,R}>2\pi
\end{array}\right.
\ee
and the thermodynamic entropy is
\be
S = (1-\beta_L \p_{\beta_L} - \beta_R \p_{\beta_R})\log Z \sim \frac{\pi^2 c}{3}\left( \frac{1}{\beta_L} + \frac{1}{\beta_R}\right) \ .
\ee
Legendre transforming to the microcanonical ensemble, this implies the Cardy behavior
\be\label{doubles}
S(E_L, E_R) \sim 2\pi\sqrt{\frac{c}{6}E_L} + 2\pi\sqrt{\frac{c}{6}E_R} \ , \quad\quad (E_{L,R} > \c{24}) \ .
\ee
It is straightforward to prove using the method of appendix \ref{app:micro} that this Legendre transform is an accurate calculation of the microscopic density of states.
For states outside the range (\ref{doubles}), we can again
only give an upper bound. The condition
\be
\rho(E_L,E_R) e^{-\beta_LE_L -\beta_R E_R} \leq Z(\beta_L, \beta_R) 
\ee
gives the constraint:
\be\label{seeall}
S(E_L, E_R) \lesssim \frac{\pi c}{6} + 2\pi(E_L+E_R) \quad\quad (\mbox{all \ } E_{L,R})
\ee
\be\label{seemixed}
S(E_L,E_R) \lesssim \frac{\pi c}{12} + 2 \pi E_L + 2 \pi \sqrt{\frac{c}{6} E_R} \quad\quad (E_R > \cb \ , \mbox{all\ } E_L)
\ee
and similarly for $L \leftrightarrow R$.

\subsection{Mixed temperature regime}\label{ss:mixed}
Let us now turn to the regime where one temperature is
high and the other is low. The situation here is
more complicated, but
we will derive universal behavior for part of this range. 
For this purpose however (\ref{canbound}) is no longer
good enough, and we need to replace it by something stronger.
To this end it is useful to change the definition of
`light' and `heavy' states 
\be\label{lightheavymixed}
L: \ E_L<0 \mbox{\ or\ } E_R<0 \ , \qquad H: \ E_R>0 \mbox{\ and\ } E_L>0 \ .
\ee
The partition function is given by
\be
Z(\beta_L, \beta_R) = \ZB{L} + \ZB{H}  \ 
\ee
where the notation $\ZB{\cdots}$ means the contribution to $Z(\beta_L,\beta_R)$ from the range
specified in (\ref{lightheavymixed}).
Our strategy is then the same as in section~\ref{s:canonical}:
We first impose constraints on the growth of the light states
in such a way that their total contribution to leading
order is still given by the vacuum contribution, and
then check if this is enough to ensure that the
full phase diagram is universal, or if the heavy states 
can make non-universal contributions.
For the first step we want to make sure that
\be\label{LoverLp}
\ZB{L} \approx \exp\left[ \c{24}(\beta_L + \beta_R)\right] \ 
\ee
for $\beta_L\beta_R>4\pi^2$. This is the case if the growth of the light states is bounded by
\be\label{mxbnd}
\rho(E_L,E_R) \lesssim \exp\left[4\pi \sqrt{(E_L + \c{24})(E_R+\c{24})}\right]  \qquad (E_L<0 \mbox{\ or\ } E_R<0) \ .
\ee
To see this, we require $\rho(E_L,E_R) \leq e^{\c{24}(\beta_L+\beta_R) + \beta_L E_L + \beta_R E_R}$ and then optimize over $\beta_{L,R}$ in the range $\beta_L\beta_R>4\pi^2$.  
This guarantees that the light states give a universal contribution
to the free energy. Next we want to check
if $\ZB{H}$ is subleading in this range.
For concreteness let us take $\beta_L > \beta_R' > 2\pi$. The
other case can be obtained by exchanging $L\leftrightarrow R$. 
We then need to bound $\ZB{H}$, and optimally we would hope
to find the analogue of (\ref{Hbound}), which would
ensure that the heavy states never dominate in this
regime. Assuming only (\ref{mxbnd}), we show in appendix~\ref{app:mixed} the
slightly weaker result
\be\label{HeavyBound}
\ZB{H} \lesssim \exp\left[\frac{\pi c}{12} + \c{24}\beta_R'\right] \ .
\ee
Unlike the case of zero angular potential, this is not enough to derive a universal free energy for all temperatures, 
as it is not dominated by (\ref{LoverLp}) in the entire range we are considering. 
We do, however, find universal behavior in the range where 
$\ZB{H} \ll \exp\left[ \c{24}(\beta_L +\beta_R)\right]$, \ie for $\beta_L > 2\pi + \beta_R' - \beta_R$, in which case
indeed 
\be
Z(\beta_L, \beta_R) = \ZB{L} + \ZB{H} \approx \exp\left[ \c{24}(\beta_L + \beta_R)\right] \ .
\ee
In total we get
\be\label{mixres}
\log Z(\beta_L,\beta_R) \sim \c{24}\max(\beta_L + \beta_R,\beta_L' + \beta_R') \qquad (\beta_L,\beta_R) \notin \cS_2\ .
\ee
The sliver around $\beta_L\beta_R=4\pi^2$, 
\be\label{sliver}
\cS_2 = \{ \beta_L < 2\pi + \beta_R' - \beta_R, \beta_R<2\pi \} + L \leftrightarrow R + \beta_{L,R} \leftrightarrow \beta'_{L,R} \ ,
\ee
 is the regime where the heavy states 
can contribute so that the free energy is not fixed so far.  This extends the previous results to the region labeled `second iteration' in figure \ref{fig:slivers}.

Turning to the microscopic spectrum, by the usual argument we obtain
\be
S(E_L,E_R) \sim S_{Cardy}(E_L,E_R) \quad (0<E_R<\cb , E_L>g_0(E_R))
\ee
\be
g_0(E_R) \equiv E_R - \cb + \frac{c^2}{576 E_R} + \frac{\cb - E_R}{\sqrt{6E_R/c}} \ .
\ee
We can also place an upper bound on a certain range where one energy is large and the other is small. Let $0<E_R< \cb$. In the inequality $\rho(E_L,E_R)e^{-\beta_LE_L-\beta_RE_R} < Z$, choose
\be
\beta_R = \frac{\pi\sqrt{c}}{\sqrt{6 E_R}} \  , \quad \beta_L' = 2\pi + \beta_R - \beta_R'
\ee
which falls in the regime where (\ref{mixres}) is applicable.
This implies
\be\label{seecurve}
S(E_L, E_R) \lesssim  g_1(E_R) E_L + g_2(E_R) \quad (0<E_R < \cb, E_L>0)
\ee
where 
\be
g_1(E_R) = \frac{2\pi \sqrt{cE_R}}{\sqrt{24} (\cb - E_R)+ \sqrt{cE_R} } \ , \quad
g_2(E_R) = \frac{\pi c}{12} + \frac{\pi c}{24\sqrt{6 E_R/c}}  + \pi \sqrt{\frac{c}{6}E_R} \,.
\ee
We can now perform another step in our iteration.
Although the free energy is not universal inside the sliver $\cS_2$, (\ref{HeavyBound}) still imposes an upper bound, which 
we can use to give a stronger bound on the microscopic spectrum. The modular transform of (\ref{HeavyBound}) implies 
\be
Z \lesssim \exp\left[ \frac{\pi c}{12}+ \c{24}\beta_R\right] \qquad (2\pi < \beta_R < \beta_L' < 2\pi + \beta_R - \beta_R') \ .
\ee
Requiring $\rho < Z e^{\beta_R E_R + \beta_L E_L}$ and minimizing over $\beta_L$, we find
\be
\rho(E_L,E_R) \lesssim \exp\left[ \frac{\pi c}{12} + \c{24}\beta_R + \frac{4\pi^2}{2\pi +\beta_R-\beta_R'}E_L + \beta_R E_R\right] \ ,
\ee
for any $\beta_R > 2\pi$. The optimal bound is obtained by minimizing this expression over $\beta_R$.  This involves solving a quartic equation, so this step is performed numerically.  However it is straightforward to see analytically that for $E_R=0$, this implies the asymptotic behavior
\be
\rho(E_L, 0) \lesssim \exp\left[ 2\pi \sqrt{\c{6}E_L} \right] \qquad (E_L\to \infty) \ ,
\ee
which is stronger than any of our previous bounds. When we apply this bound on the spectrum to the free energy, it reduces the size of the unknown range to a smaller sliver $\cS_3$, as shown in the `third iteration' of figure \ref{fig:slivers} where $\cS_3$ is the white region. 
The range of energies where the Cardy formula applies to the microsopic spectrum becomes very close to the line $E_LE_R=(c/24)^2$, as is shown in figure \ref{fig:phases}\textcolor{darkblue}{b}. 

One can of course continue with this procedure iteratively. We conjecture that the sliver would collapse onto the line $\beta_L\beta_R=4\pi^2$. That is, we expect (but have not shown) that the leading free energy is universal everywhere away from the self-dual line,
\be\label{Funiversal}
\log Z(\beta_L,\beta_R) \sim \c{24}\max(\beta_L + \beta_R,\beta_L' + \beta_R') \qquad (\beta_L\beta_R\neq4\pi^2)\ .
\ee
In this case, using $\rho(E_L, E_R) \leq Z e^{\beta_L  E_L + \beta_R E_R}$ with (\ref{Funiversal}) and optimizing the bound over $\beta_{L,R}$ implies 
\be\label{nonCardyBound}
S(E_L, E_R) \lesssim 4\pi \sqrt{\left(E_L + \c{24}\right)\left(E_R + \c{24}\right)} \ ,
\ee
for all $E_{L,R}  > - \c{24}$. 
Moreover, repeating the arguments in section~\ref{ss:mixspectrum}, we can transform (\ref{Funiversal}) 
to the microcanonical ensemble to get 
\be\label{CardyS}
S(E_L, E_R)  \sim S_{Cardy}(E_L, E_R) \ , \quad \mbox{\ for \ } \quad  E_LE_R > \frac{c^2}{576} \ .
\ee
The usual arguments (see appendix \ref{app:micro}) imply that this expression is accurate in the microcanonical ensemble to leading order in $1/c$.

\section{Comparison to 3d gravity}\label{s:gravity}

Black holes provide UV data about quantum gravity, such as the approximate density of states at high energy.  Since their thermodynamics is determined by the low energy effective action, this means that any UV completion of quantum gravity shares a number of universal features.  In this section we will review some of the well known universal features of 3d gravity, and show that they correspond exactly to the universal properties of 2d CFT at large $c$ derived above.

\subsection{Canonical ensemble}

Any theory of gravity+matter in AdS$_3$ has (at least) two competing phases at finite temperature: the BTZ black hole \cite{Banados:1992wn,Banados:1992gq} and a thermal gas.  The black hole action is \cite{Maldacena:1998bw}
\be
\log Z_{BH} = \frac{\pi^2 c}{6}\left( \frac{1}{\beta_L} + \frac{1}{\beta_R}\right) \ ,
\ee
where $c = 3\ell/2G_N$,  with $\ell$ the AdS radius and $G_N$ Newton's constant. The thermal gas is the same classical solution as empty AdS but in a different quantum state.  Its classical action is that of global AdS,
\be
\log Z_{therm} = \frac{c}{24}(\beta_L + \beta_R) \ .
\ee
Both of these classical solutions obey the same finite-temperature boundary condition, and in the canonical ensemble the partition function is a sum over such saddlepoints.  Therefore, $Z_{grav}(\beta) \approx e^{-I_{BTZ}} + e^{-I_{therm}} + \cdots$ with $I$ the Euclidean action, and we find
\be\label{zgrav}
\log Z_{grav}(\beta_L, \beta_R) \approx \max\left( \log Z_{BH} \ , \ \log Z_{therm}\right) \ .
\ee
There is a Hawking-Page phase transition at $\beta_L + \beta_R = \beta_L' + \beta_R'$ \cite{Maldacena:1998bw,Hawking:1982dh,Witten:1998zw}. 

In principle, other saddlepoints should also be included.  Even without matter fields, there is an infinite family of Euclidean solutions in pure gravity known as the $SL(2,Z)$ black holes.  These are obtained from the Lorentzian black hole by the analytic continuation to imaginary angular potential,
\be
\tau = \frac{i\beta_L}{2\pi} , \quad \btau = - \frac{i\beta_R}{2\pi} \ ,
\ee
followed by the $SL(2,Z)$ transformation $\tau \to \frac{a \tau + b}{c\tau + d}$.  The resulting action is
\be
\log Z = -\frac{i \pi c}{12}\left( \frac{a \tau + b}{c\tau + d} - \frac{a \btau + b}{c \btau + d}\right) \ .
\ee
Maximizing this expression over $SL(2,Z)$ images leads to an intricate Euclidean phase diagram with an infinite number of phases tessellating the upper half $\tau$-plane \cite{Dijkgraaf:2000fq, Maldacena:1998bw, Maloney:2007ud}. However, in Lorentzian signature, $\beta_{L,R}$ are real and cosmic censorship imposes $\beta_{L,R} \geq 0$. This translates under analytic continuation into
\be\label{eco}
|\mbox{Re}\, \tau | \leq \mbox{Im}\, \tau \ .
\ee
Within this range, the dominant phase is either Euclidean BTZ or thermal AdS.  In other words, when we compute the free energy for real angular potential, these are the only two dominant phases in pure gravity.  Allowing for matter fields could lead to new saddlepoints, but we do not know of any example where the new saddlepoints dominate the canonical ensemble.  

At zero angular potential, the gravity result (\ref{zgrav}) precisely agrees with our CFT result (\ref{vaconly}) for all values of the temperature.  At finite angular potential, the gravity formula was derived from CFT for all $\beta_{L,R}$ except within the sliver discussed in section \ref{ss:mixed}.  This can be viewed as a prediction that in any theory of gravity+matter, BTZ or thermal AdS is indeed the dominant saddlepoint (at least outside the sliver).

\subsection{BTZ black holes in the microcanonical ensemble}
The known phases of 3d gravity in the microcanonical ensemble are much richer.  In addition to BTZ black holes, there are other bulk solutions  with $O(c)$ entropy, including black holes localized on the internal manifold \cite{deBoer:2008fk} and multicenter solutions \cite{Bena:2011zw}.  Within certain parameter ranges, these can have entropy greater than BTZ and thus dominate the microcanonical ensemble.  Before turning to these more exotic solutions let us compare the spectrum and entropy of the BTZ black hole to our CFT results.  BTZ black holes have energies
\be
E_{L,R} = \frac{\pi^2 c}{6\beta_{L,R}^2} \ , 
\ee
and entropy given by the Cardy formula
\be\label{sbhc}
S_{BH}(E_L,E_R) = S_{Cardy}(E_L,E_R) \ .
\ee
They exist for all $E_{L,R} \geq 0$. 

To compare to CFT, first consider the case of zero angular momentum $E_L=E_R=E/2$. The black holes exist and have Cardy entropy for $E\geq 0$, but in the CFT we only derived the Cardy entropy for $E > \c{12}$ (see section \ref{s:canonical}). In fact this is perfectly consistent: the black holes with $0 <  E < \c{12}$ are unstable in the canonical ensemble. These unstable black holes eventually tunnel into the gas phase. Therefore within this range the black holes are subleading saddlepoints, much like the subleading saddles in CFT discussed in section \ref{ss:cftsubsaddles}. There we argued that, generically (assuming no delicate cancellations), the subleading saddle in CFT gives a reliable contribution to the microscopic density of states; this contribution corresponds exactly to the unstable black holes.

The situation at finite real angular potential is similar. In the regime where we found a universal CFT entropy given by the Cardy formula, it agrees with the entropy of rotating BTZ (\ref{sbhc}).  Outside the universal regime, we derived an upper bound on the CFT density of states which is satisfied by (\ref{sbhc}). Subleading saddlepoints in the CFT with rotation were not discussed, but are easily seen to correspond to unstable black holes with $\beta_L \beta_R> 4\pi^2$. 

\subsection{Enigmatic phases in the microcanonical ensemble}\label{ss:enigmaticgravity}
As mentioned above, there are known solutions in 3d gravity with entropy greater than that of BTZ at the same energies,
\be
S_{enigma}(E_L, E_R) > S_{Cardy}(E_L, E_R) \ .
\ee
The examples we will consider are the $S^2$-localized black holes in \cite{deBoer:2008fk} and the moulting black holes in \cite{Bena:2011zw}. These are similar to the enigmatic phases discussed in \cite{Gauntlett:2004wh,Denef:2007vg} so we adopt this terminology.  

We will see that the enigma saddlepoints fit nicely with our CFT results.  They fall in the intermediate range $0\leq E_{L,R} \leq \c{24}$, where we found that the CFT entropy is not universal but obeys
\be\label{cftrange}
S_{Cardy}(E_L,E_R) \leq S_{CFT}(E_L,E_R) \leq \frac{c\pi}{6}+2\pi (E_L+E_R)\ .
\ee
The upper bound holds universally, while the lower bound holds provided we assume that subleading saddlepoints are not cancelled.  The upper bound is simply the statement that these states never dominate the canonical ensemble.

The relevant solutions in \cite{deBoer:2008fk} are BPS solutions of M-theory compactified on $S^1 \times $CY$_3$.  In the decoupling limit, the 5d geometry is asymptotically an $S^2$ fiber over $AdS_3$. From a higher-dimensional perspective the twisting of the fiber is proportional to angular momentum; from the  3d gravity or dual CFT point of view, twisting corresponds to $SU(2)_R$ charge.
At high energies, the highest-entropy BPS solution with these asymptotics is an uncharged extremal BTZ$\times S^2$ with energies $(E_L, 0)$ and entropy given by the Cardy formula.  However there is another solution in which the black hole is localized on the $S^2$. This solution carries $SU(2)_R$ charge but can nonetheless dominate over uncharged BTZ.  (Multicenter localized black holes, including some with zero $SU(2)_R$ charge, are also discussed in \cite{deBoer:2008fk}  but these have lower entropy.) The localized solution exists for $-\c{24}<E_L<\frac{9c}{128}$ and at the BTZ threshold $E_L=E_R=0$ it has entropy
\be\label{mswen}
S_{enigma} = \frac{\pi c}{18\sqrt{3}} \ .
\ee
The scaling of (\ref{mswen}) with $c$ indicates that this solution has more entropy than BTZ in some range just above the threshold.  The transition point is \cite{deBoer:2008fk}
\be
E_L^{crit} \approx 0.046 \c{24} \ .
\ee
Thus the microscopic entropy is greater than the Cardy formula for $0<E_L<E_L^{crit}$, and falls within our CFT bounds (\ref{cftrange}). As expected from CFT, the localized black hole never dominates the canonical ensemble. 

As a second example we turn to the two-center solution of IIB supergravity compactified on $T_4$ constructed in \cite{Bena:2011zw}. This solution, which is described as a BMPV black hole surrounded by a supertube, has near horizon geometry $AdS_3 \times S^3$ so our results should apply. The entropy of the new solution (spectral flowed to the NS sector) is
\be
S(E_L) = 2\pi\left(\sqrt{\c{6}} - \sqrt{ \c{8}-E_L }\right)\sqrt{E_L + \c{24}} \ ,
\ee
and it exists for $-\c{24} < E_L < \c{24}$.  This dominates over the Cardy entropy in a small window above $E_L = 0$ up to the critical value
\be
E_L^{crit} \approx 0.019 \c{24} \ .
\ee
Once again these states obey (\ref{cftrange}) and never dominate the canonical ensemble.

 The gravity examples that we have considered here are supersymmetric, but our CFT results suggest that entropy above the Cardy value at intermediate energies is a generic feature of large $c$ CFTs. Since we did not find a universal answer for $S_{CFT}$ in this range, we cannot check the explicit formula for $S_{enigma}$ from CFT beyond confirming that it obeys the bounds.  Indeed, we expect that $S_{enigma}$ depends on the specific microscopic theory, and in particular it may depend on the coupling constant.

\section{Example: Symmetric orbifolds}\label{s:orbifold}

So far our discussion has been general, as it applies to any unitary, modular invariant CFT with large $c$ and sparse low-lying spectrum.  We now turn to a specific class of examples, symmetric orbifold CFTs, to illustrate how these theories fit into our general picture. 
Symmetric orbifold CFTs have been studied extensively in the context of the D1-D5 system. They were used in the original computation of \cite{Strominger:1996sh}, and underlie many of the more recent successful precision tests of black hole microstate counting in string theory summarized for example in \cite{Sen:2007qy,Sen:2012cj}. 
We will show that all symmetric orbifold theories have
the universal free energy (\ref{Funiversal}), which of course implies
that they satisfy the constraints on the spectrum (\ref{CardyS}) and (\ref{nonCardyBound}). 
In fact symmetric orbifolds
saturate the bound (\ref{nonCardyBound}). This shows that in a sense
they are most dense theories that are still compatible
with the universal free energy (\ref{Funiversal}).

 Starting with any `seed' theory $\C$, the symmetric orbifold $\C^N/S_N$ consists of $N$ copies of the original theory, orbifolded by the permutation group.  If we take the seed theory to be the sigma model with target space $M_4$, where $M_4 = K3$ or $T^4$, then the symmetric orbifold CFT is holographically dual to IIB string theory on $AdS_3 \times S^3 \times M_4$.  The seed theory has central charge $c_1 = 6$ and the orbifold has $c = N c_1$.  The orbifold theory itself is the weak coupling limit and does not have a good geometrical description, but in principle we can turn on exactly marginal deformations in the CFT to reach a point in moduli space with a semiclassical gravity description.

The spectrum of the D1-D5 CFT depends on the moduli, so the spectrum of the symmetric orbifold need not match the spectrum of supergravity, while certain supersymmetric quantities (such as the elliptic genus) are protected and can be successfully matched on the two sides of the duality. Relatively little is known about the non-supersymmetric features of the CFT at strong coupling, except what is fixed entirely by symmetry or has been deduced from the gravity picture. On the other hand, the results of sections \ref{s:canonical} - \ref{s:angularpot} do not require supersymmetry, and apply to the D1-D5 CFT in the gravity limit (if our assumptions about the light spectrum are satisfied) as well as at the orbifold point.

In this section we will compute the density of states at the orbifold point, for an arbitrary seed theory. We show that it satisfies our assumptions about the light spectrum (\ref{introasum}, \ref{introamix}), and confirm that the heavy spectrum is consistent with our results.  
Symmetric orbifolds also saturate the upper bound (\ref{nonCardyBound}) in the enigmatic range $0<E<\c{12}$, demonstrating that this bound is optimal.

Some of these results have previously been derived using the long string description of the D1-D5 system, but the explicit orbifold CFT computation is instructive to make precise exactly when the long string picture is reliable. The result in section \ref{ss:orbs} for the spectrum of light states appears to be new.

\subsection{Partition function}
The partition function of a symmetric orbifold is determined by the seed theory.  Let us choose a seed theory $\C$ and denote its partition function by
\be
Z_1 = \Tr q^{L_0 - \frac{c_1}{24}}\bq^{\overline{L}_0 -  \frac{c_1}{24}} = q^{-c_1/24}\bq^{-c_1/24}\sum_{h, \bh \in I} d_1(h, \bh) q^{h} \bq^{\bh} \ ,
\ee
where the sum is over a discrete spectrum $I$ of conformal dimensions, $h,\bh \geq 0$.  The Euclidean notation is related to the Lorentzian notation in the rest of the paper via
\be
q = e^{-\beta_L}, \quad \bq = e^{-\beta_R} 
\ee
i.e., $q = e^{2\pi i \tau}, \bq = e^{-2\pi i \btau},  \tau = \frac{i \beta_L}{2\pi}  , \btau = - \frac{i \beta_R}{2\pi}$. 
The partition function $Z_N$ of the symmetric orbifold $\C^N/S_N$,
\be
Z_N = q^{-c_1 N/24}\bq^{-c_1 N/24}\sum_{h, \bh} d_N(h, \bh) q^{h} \bq^{\bh} \ ,
\ee
is obtained as usual by projecting out states that are not invariant
under permutations, and introducing twisted sectors.
In practice it can be extracted from its generating function,
for which a relatively simple expression exists \cite{Keller:2011xi,Dijkgraaf:1996xw}: 
\be\label{calz}
\Z \equiv \sum_{N\geq 0} p^N Z_N = \prod_{n>0} \prod_{h,\bh \in I}(1-p^n q^{(h-c_1/24)/n}\bq^{(\bh-c_1/24)/n})^{-d_1(h,\bh)\delta^{(n)}_{h-\bh}} \ .
\ee
Here roughly speaking $n$ corresponds to the length of the twisted
sectors, and 
\be
\delta^{(n)}_{h-\bh} =\left\{\begin{array}{cc} 1 :& h-\bh = 0\mod n \\
0 :& \textrm{else} \end{array}\right.
\ee
projects out states of non-integer spin. In \cite{Keller:2011xi} this expression was used to show that 
the free energy of large-$N$ symmetric orbifolds has universal thermodynamic behavior for $\tau$ in the upper half complex plane.  
In appendix \ref{app:rfinite} we repeat this argument for real angular potential to prove
\be\label{symph}
\log Z_N = \c{24}\max\left( \beta_L + \beta_R, \beta_L' + \beta_R'\right)  +O(1) \ ,
\ee
for all $\beta_{L,R} > 0$, where throughout this section $c = c_1 N$. This is somewhat stronger than (\ref{mixres}) derived in section \ref{ss:mixed}, because it also applies in the sliver $\cS$.

\subsection{Spectrum}\label{ss:orbs}
Let us now discuss the spectrum of the theory. 
We established above that the free energy satisfies (\ref{Funiversal}),
from which it follows that the bound (\ref{nonCardyBound}) is satisfied.
In appendix \ref{app:saturated}, we prove that this bound is actually saturated,
\be\label{smain}
S(E_L, E_R) \sim 4\pi \sqrt{(E_L + \c{24})(E_R + \c{24})} \quad \mbox{\ for\ } \quad E_LE_R < \frac{c^2}{576} \ .
\ee
Together with (\ref{CardyS}) this fixes the spectrum of symmetric orbifold
theories completely, and shows that it is completely universal, \ie
depends only on the central charge.
A detailed derivation of (\ref{smain}) can be found in the appendix. The general idea is that we are counting the excitations of $N$ strings that can join into longer strings.  Long strings have Cardy entropy in the range (\ref{CardyS}). For a given $(E_L,E_R)$, the entropy (\ref{smain}) comes from the sector with $M$ short strings and one long string (made of $N-M$ short ones), maximized over $M\leq N$. 

The entropy at energy $E = E_L + E_R$ is dominated by $E_{L,R} = E/2$, which gives
\be
S(E) \sim \frac{\pi c}{6} + 2 \pi E \quad\quad (0 < E < \c{12}) \ .
\ee
Thus the symmetric orbifold saturates our upper bound in (\ref{specup}) in the enigmatic regime. Pure gravity, on the other hand, saturates the lower bound, while known UV-complete theories of 3d gravity+matter appear to fall in between, as discussed in section \ref{ss:enigmaticgravity}. This implies that going to strong coupling in CFT lifts some of the enigmatic states (similar conclusions were reached in \cite{deBoer:2008fk,Bena:2011zw}).

\bigskip

\textbf{Acknowledgments} We thank Dionysios Anninos, Daniel Friedan, Matthias Gaberdiel, Alex Maloney, Don Marolf, Greg Moore, Hirosi Ooguri, Eric Perlmutter, Andrea Puhm, and Andy Strominger for useful discussions. TH is supported by the National
Science Foundation under Grant No. NSF PHY11-25915. CAK is supported by the Rutgers
New High Energy Theory Center and by U.S. DOE Grants No.~DOE-SC0010008,
DOE-ARRA-SC0003883 and DOE-DE-SC0007897. CAK thanks the
Harvard University High Energy Theory Group for hospitality. BS is supported in part by a Dominic Orr Graduate Fellowship and by U.S. DOE Grant No.~DE-SC0011632. BS would like to thank the Kavli Institute For Theoretical Physics for hospitality.

\appendix

\section{Density of states in the microcanonical ensemble}\label{app:micro}

The exact density of states is a sum of delta functions, so to make equations like $\rho(E) \approx e^{S(E)}$ precise requires averaging over an interval.  For this we introduce
\be
n_{\ec,\delta} = N_{states}(\ct\ec-\delta < E < \ct\ec + \delta) \ ,
\ee
which counts the number of states in an interval around some energy. For the exponential dependence, the distinction between number $n_{\ec,\delta}$ and number density $\rho$ is not important. We will take $\ec$ fixed and independent of $c$. The size of the interval $\delta$ on the other hand needs to increase with $c$. Choosing the correct scaling with $c$ is actually crucial. It turns out that we need it to scale as $\delta \sim c^{\alpha}$ with  $\frac{1}{2}<\alpha<1$. With this scaling we can show that
\bear\label{nval}
\log n_{\ec,\delta} \leq \frac{\pi c}{6}(1+\ec) + O(c^\alpha) &:& 0<\ec <1\\
\log n_{\ec,\delta} =\frac{\pi c \sqrt{\ec}}{3} + O(c^\alpha) &:& \ec > 1 \label{nval2}\ ,
\eear
that is, we show that (\ref{sereg}) and (\ref{eupi}) indeed hold microscopically. This already shows why we needed to pick $\alpha<1$, since otherwise the density would obtain corrections of order $c$ or bigger. To prove (\ref{nval}) it will be useful to decompose the heavy spectrum
$H$ into 
\be
H_1 =\left\{\epsilon<E<\frac{c\ec}{12}-\delta\right\}\ ,\quad 
H_2 =\left\{\frac{c\ec}{12}-\delta\leq E<\frac{c\ec}{12}+\delta\right\}\ ,\quad  
H_3 =\left\{\frac{c\ec}{12}+\delta\leq E\right\}\ .
\ee
Let us first construct the upper bound.
For $\beta <2\pi$ we have 
\begin{multline}\label{nabove}
\beta' \ct = \log Z(\beta) +O(1) = \log \ZB{H} +O(1) \\ 
\geq \log\ZB{H_2} + O(1) 
 \geq \log\left( n_{\ec,\delta}e^{-\beta(\ct\ec+\delta)}\right) +O(1)
\end{multline}
so that
\be
\log n_{\ec,\delta} \leq \frac{\pi^2 c}{3\beta} + \beta (\ct\ec+\delta) + O(1)  \ .
\ee
We can optimize this bound by picking $\beta = 2\pi/\sqrt{\ec}$
if $\ec >1$, or $\beta =2\pi$ if $\ec<1$. Using $\delta = O(c^\alpha)$ it follows that 
\bea\label{upperb}
\log n_{\ec,\delta} &\leq& \frac{\pi c \sqrt{\ec}}{3} + O(c^\alpha) \quad\quad (\ec >1) \ , \\ 
\log n_{\ec,\delta} &\leq& \frac{\pi c}{6}(1+\ec) +O(c^\alpha) \quad\quad (\ec <1) \ .
\label{upperb2}
\eea
To derive (\ref{nval2}), we must show that (\ref{upperb}) is saturated.  
The idea is again to pick a specific $\beta$ so that the main contribution
to $\ZB{H}$ comes from the states at $\ec$.
Setting $\beta = 2\pi/\sqrt{\ec}$,
we first want to show that 
\be\label{HH2}
\log \ZB{H} = \log \ZB{H_2} + O(1)\ .
\ee
To this end we estimate
\be\label{H3}
\log \ZB{H_3} \leq  \frac{\pi c}{3}(\sqrt{\ec+12\delta/c}-\frac{1}{2}\frac{\ec+12\delta/c}{\sqrt{\ec}})
+O(\log c)
= \frac{\pi c\sqrt{\ec}}{6} - \frac{6\pi \delta^2}{ \ec^{3/2}c} + o(c^{2\alpha-1})\ ,
\ee
where in the first equality we have used that the total
sum differs from its maximal summand only by
a polynomial prefactor. Since the first subleading term
comes with a negative sign and grows as $c^{2\alpha-1}$,
it follows from $\ZB{H}= \frac{\pi c\sqrt{\ec}}{6}+O(1)$
that
\be
\ZB{H_3}/\ZB{H} \rightarrow 0\ .
\ee
We can show a similar result for $\ZB{H_1}$: Here we
split $H_1$ into $H_4 = \left\{\epsilon<E<1\right\}$
and $H_5=\left\{1<E<\ct\ec-\delta\right\}$.
The contribution from $H_4$ we can estimate
using (\ref{upperb2}) as
\be
\log\ZB{H_4} \leq \frac{\pi c}{6}\sqrt{\ec}(1-(1-\ec^{-1/2})^2) +O(\log c)\ ,
\ee
and the contribution from from $H_5$ using
(\ref{upperb}), which gives (\ref{H3})
but with $-\delta$ instead of $\delta$.
Combining these three estimates, (\ref{HH2}) follows, and then
we can use
\be
\ZB{H_2} \leq \log n_{\ec,\delta}e^{-\frac {c\pi}{6\sqrt{\ec}}(\ec-12\delta/c)}
\ee
to obtain the lower bound that leads to (\ref{nval2}).

\section{Mixed temperature calculations}\label{app:mixed}
This appendix contains the details of the calculation discussed in section \ref{ss:mixed}. 
We assume $\beta_L>2\pi>\beta_R$ and $\beta_L \geq \beta_R'$,
which in particular implies $\beta_L+\beta_R \geq 4\pi$.
To establish (\ref{HeavyBound}), we need to bound 
$\ZB{H}$. We decompose it into 4 terms
\bea
T_1 &=& \ZB{\c{24}<E_L, \c{24}<E_R}\\
T_2 &=& \ZB{0< E_L < \c{24}, \c{24}<E_R}\\
T_3 &=& \ZB{\c{24}<E_L ,0<E_R <\c{24}}\\
T_4 &=& \ZB{0<E_L < \c{24},0<E_R<\c{24}}\ ,
\eea
and then apply the various bounds (\ref{doubles}), (\ref{seeall}) and (\ref{seemixed}). 
For $T_1$ we use (\ref{doubles}),
\bea
T_1 &\lesssim& \int_{\c{24}}^\infty dE_L \int_{\c{24}}^\infty dE_R \exp\left[ 2\pi \sqrt{\frac{c}{6}E_L} + 2\pi \sqrt{\c{6}E_R} - \beta_L E_L - \beta_R E_R\right]\notag \\
&\approx& \exp\left[\c{24}(4\pi-\beta_L + \beta_R')\right] \ll \exp\left[ \c{24}(\beta_L + \beta_R)\right] \ ,
\eea
the leading contribution coming from $E_L = \c{24}, E_R =\frac{\pi^2 c}{6\beta_R^2}>\c{24}$.
The term $T_2$ is in the range where the bound (\ref{seemixed}) applies. Thus
\bea
T_2 &\lesssim& \int_{\c{24}}^\infty dE_R \int_0^{\c{24}} dE_L \exp\left[ \frac{\pi c}{12} + 2 \pi E_L + 2 \pi \sqrt{\c{6}E_R} - \beta_L E_L - \beta_R E_R\right]\notag\\
&\approx& e^{\pi c/12} \int_{\c{24}}^\infty dE_R \exp\left[ 2\pi\sqrt{\c{6}E_R} - \beta_R E_R\right]\notag\\
&\approx & \exp\left[ \frac{\pi c}{12} + \c{24}\beta_R'\right]\ .
\eea
The dominant term here comes from $E_L = 0, E_R =\frac{\pi^2 c}{6\beta_R^2}$. 
For $T_3$ we apply the flipped version of (\ref{seemixed}),
\bea
T_3 &\lesssim& \int_{\c{24}}^\infty dE_L \int_0^{\c{24}} dE_R \exp\left[ \frac{\pi c}{12} + 2 \pi E_R + 2 \pi \sqrt{\c{6}E_L} - \beta_L E_L - \beta_R E_R\right]\notag\\
&\approx& e^{\c{24}(4\pi-\beta_R)} \int_{\c{24}}^\infty dE_L \exp\left[ 2\pi\sqrt{\c{6}E_L} - \beta_L E_L\right]\notag\\
&\approx & \exp\left[ \c{24}(8\pi-\beta_L - \beta_R)\right] \ll \exp\left[\c{24}(\beta_L + \beta_R)\right]\ .
\eea
Finally for $T_4$ we use (\ref{seeall}) to get
\bea
T_4 &\lesssim& \int_0^{\c{24}} dE_R \int_0^{\c{24}} dE_L \exp\left[\frac{\pi c}{6} + 2 \pi(E_L + E_R) -\beta_L E_L - \beta_R E_R\right] \notag\\
&\approx& \exp\left[ \frac{\pi c}{4} - \c{24}\beta_R\right] \ll T_2 \ ,
\eea
where the dominant contribution comes from $E_L=0$ and $E_R=\c{24}$.
In total we have shown
\be
\ZB{H}  \lesssim \exp\left[ \frac{\pi c}{12} + \c{24}\beta_R'\right] \ .
\ee

\section{Symmetric orbifold calculations}
\subsection{Free energy}\label{app:rfinite}
In this appendix we use (\ref{calz}) to derive the large-$N$ phases of the symmetric orbifold at real angular potential claimed in (\ref{symph}). The argument parallels the Euclidean discussion in \cite{Keller:2011xi} so we will be brief.  Suppose $\beta_L > \beta_R'$, so the first term in (\ref{symph}) dominates.  Define the remainder
\be
R_N = \log\left( Z_N e^{-\c{24}(\beta_L + \beta_R)}\right) \ ,
\ee    
which gives the contribution to the free energy of all the states
other than the vacuum.
We will prove that this is a subleading contribution by showing that $R_\infty$ is finite.   
Using (\ref{calz}), it is straightforward to derive (see \cite{deBoer:1998us}
and in particular section 2.2.3 and appendix A.2 of \cite{Keller:2011xi})
\be\label{Rinfty}
R_\infty =  \sum_{n>0}\sum_{k>0}\sum_{h,\bh \in I}\!\!' \frac{1}{k}d_1(h,\bh)\delta^{(n)}_{h-\bh}q^{kh/n + k\frac{c_1}{24}(n-1/n)}\bq^{k\bh/n + k\frac{c_1}{24}(n-1/n)}
\ee
where the primed sum indicates that we skip the term with $n=1, h=\bh = 0$. Every term is positive so in checking convergence we can ignore the delta and exchange sums at will. The $n$th term for $n>1$ is then simply 
\be\label{nterm}
\sum_{k>0}\frac{1}{k} \exp\left[-\frac{c_1 k n}{24}(\beta_L+\beta_R)\right] Z_1\left(\frac{k}{n}\beta_L, \frac{k}{n}\beta_R\right) \ .
\ee
To proceed we will bound the seed partition function $Z_1$ that appears in this expression
by 
\be\label{Z1estimate}
Z_1(\beta_L, \beta_R) \leq p(\beta_L,\beta_R) e^{\frac{c_1}{24}(\beta_L+\beta_R)}e^{\frac{c_1}{12}(\beta_L' + \beta_R')} \ , 
\ee
where $p(\beta_L,\beta_R)$ grows at most polynomially.
To see this note that the standard Cardy formula tells us that
for all $h$ and $\bar h$
\be\label{Cardybound}
\rho(h+\bar h) \leq N e^{2\pi \sqrt{c_1(h+\bar h)/3}}\ 
\ee
for some constant $N$. (This follows from the fact that (\ref{Cardybound}) holds
asymptotically for large $h+\bar h$, so we simply choose $N$
large enough so that it holds everywhere.) It follows that
\begin{multline}
Z_1(\beta_L,\beta_R) = e^{\frac{c_1}{24}(\beta_L+\beta_R)}\int dh d\bar h \rho(h,\bar h) e^{-\beta_L h}e^{-\beta_R \bar h}
\\
\leq N e^{\frac{c_1}{24}(\beta_L+\beta_R)} \int dh d\bar h e^{2\pi\sqrt{ch/3}-\beta_Lh}e^{2\pi\sqrt{c\bar{h}/3}-\beta_R\bar{h}}
\leq p(\beta_L,\beta_R) e^{\frac{c_1}{24}(\beta_L+\beta_R)}e^{\frac{c_1}{12}(\beta_L' + \beta_R')}
\end{multline}
where we have used $\rho(h,\bar h) \leq \rho(h+\bar h)$.
Plugging this into (\ref{nterm}) we can bound the exponential factors in the terms for $k>1,n>1$ by
\be\label{knbound}
e^{-\frac{n k c_1}{24}\left(\beta_L+\beta_R - \frac{1}{n^2}(\beta_L + \beta_R)- \frac{2}{k^2}(\beta_L' + \beta_R')\right)} 
\leq e^{-\frac{n k c_1}{24}\left(\frac{3}{4}(\beta_L+\beta_R) -  \frac{1}{2}(\beta_L' + \beta_R')\right)} \ .
\ee
Since by assumption $\beta_L + \beta_R > \beta_L' + \beta_R'$ the double sum over $k >1,n>1$ converges. 
The sum over $n=1,k>1$ converges since (\ref{Rinfty}) excludes the
vacuum for $n=1$, so that the exponent of the first factor in (\ref{Z1estimate})
is given by the lowest state of the theory instead.
The sum for $k=1, n>1$ converges because for 
$n$ large enough we can estimate
\be
Z\left(\frac{\beta_L}{n},\frac{\beta_R}{n}\right)=
Z(n\beta_L',n\beta_R') \leq K e^{\frac{nc_1}{24}(\beta_L'+\beta_R')}
\ee
where we can use the last inequality if $n$ is large enough so 
that $n\beta_L', n\beta_R' >2\pi$. Convergence then follows from $\beta_L + \beta_R > \beta_L' + \beta_R'$.
It follows that when $\beta_L > \beta_R'$, the free energy is indeed given
only by the vacuum contribution $\c{24}(\beta_L + \beta_R)$, and by modular invariance we obtain (\ref{symph}).

\subsection{Spectrum}\label{app:saturated}
We now derive the low-energy density of states (\ref{smain}). We have already argued that this is an upper bound, so the strategy is to find a contribution saturating this bound.  For this we will use the fact that the generating function (\ref{calz}) can be reorganized as \cite{Dijkgraaf:1996xw,Bantay:2000eq}
\be\label{zhecke}
\Z = \exp\left( \sum_{L>0} \frac{p^L}{L}\THecke_L Z_1\right) \ ,
\ee
where $\THecke_L$ is the (unnormalized) Hecke operator.  The definition of $\THecke_L$ can be found in \cite{Keller:2011xi}, but for our purposes we just need one basic fact: If $Z_1$ is a modular-invariant partition function with positive coefficients $d_1(h,\bh) > 0$, then $\THecke_LZ_1$ is also modular invariant, and can be expanded as
\be
\THecke_L Z_1 = q^{-c_1 L/24}\bq^{-c_1 L/24}\sum_{h,\bh} \dT{L}(h,\bh) q^h \bq^{\bh}
\ee
with non-negative weights $h,\bh \geq 0$ and positive coefficients $\dT{L} > 0$.

To leading order at large $N$, the degeneracy of states in the symmetric orbifold $d_N$ can be extracted from (\ref{zhecke}) by a minor extension of the argument in section 2.2.1 of \cite{Keller:2011xi}. Let
\be
\tilde{p} = p (q \bq)^{-c_1/24} \ .
\ee
Separating the contribution from the ground states in each sector,
\bea
\Z &=& \exp\left[ \sum_{L>0}\frac{\tilde{p}^L}{L} +\sum_{L>0}\frac{\tilde{p}^L}{L}\sum_{h,\bh>0}\dT{L}(h,\bh)q^{h}\bq^{\bh} \right]\\
&=& \left( \sum_{K\geq 0}\tilde{p}^K\right)\left[1+\sum_{L>0}\frac{\tilde{p}^L}{L}\sum_{h,\bh>0}\dT{L}(h,\bh)q^{h}\bq^{\bh}+\cdots\right] \ .
\eea
The corrections indicated by dots come with positive coefficients, so if we ignore the corrections then the coefficient of $\tilde{p}^N$ gives a lower bound on the orbifold degeneracy:
\be\label{dnlower}
d_N(h,\bh) \geq \sum_{L=1}^N \frac{1}{L}\dT{L}(h,\bh) \ .
\ee
In the effective string language, this equation has a simple interpretation. We are counting the degeneracy at level $(h,\bh)$ of  $N$ strings that are allowed to join into longer strings.  The $L$th term in (\ref{dnlower}) is the degeneracy in the sector with one long string and $N-L$ short strings.

Suppose for a moment that the Cardy formula applies to $\THecke_L Z_1$, so\footnote{The Cardy formula applies to the density of states, not necessarily to the degeneracy at a particular level. To be precise, in these expressions we should average $d_N$ and $\dT{L}$ over a range $(h \pm \delta, \bh \pm \delta)$ as in appendix \ref{app:micro}. We will not write this explicitly but it does not change the final answer.}
\be\label{dlcardy}
\dT{L}(h,\bh) \approx \exp\left[ 2\pi \sqrt{\frac{c_1 L}{6}(h - \frac{c_1 L}{24})} + 2\pi \sqrt{\frac{c_1 L}{6}(\bh - \frac{c_1 L}{24})} \right]\ .
\ee
The maximum in (\ref{dnlower}) occurs at
\be\label{myL}
L = \frac{24 h \bh}{c_1(h + \bh)} \ ,
\ee
which as long as $L\leq N$ would give
\be\label{dnb}
d_N(h,\bh) \gtrsim \exp\left[4\pi \sqrt{h\bh}\right] \quad \mbox{\ for\ }\quad \frac{h\bh}{h+\bh} \leq \c{24}\ .
\ee
To confirm that the argument given is reliable, we must show that the Cardy behavior (\ref{dlcardy}) holds for (\ref{myL}). Note that  
\be
\dT{L}(h,\bh) \leq L d_L(h,\bh) \ , 
\ee
\ie to leading order the $L$th Hecke transform does not have
more states than the $L$th symmetric orbifold.
It is thus straightforward to show using (\ref{symph}) that
it too has the universal free energy behavior
\be
\log \THecke_L Z_1 \sim \frac{c_1 L}{24}\max\left( \beta_L + \beta_R,\  \beta_L' + \beta_R'\right) 
\ee
as $L \to \infty$. Thus the Cardy formula (\ref{dlcardy}) applies when $E_L E_R > (c_1 L)^2/576$, \ie
\be
 \frac{h \bh}{h + \bh} \geq \frac{c_1 L}{24}\ .
\ee
The choice (\ref{myL}) falls at the edge of this range, so the bound (\ref{dnb}) is indeed valid.  Translating to energies $E_L = h - \c{24}$, $E_R = \bh - \c{24}$, (\ref{dnb}) implies that (\ref{nonCardyBound}) is saturated, which implies (\ref{smain}). Finally if $L>N$, then $\dT{N}$ provides the optimal bound, 
\be\label{dnc}
d_N(h,\bh) \gtrsim \exp\left[2\pi \sqrt{\c{6}(h - \c{24})} + 2\pi \sqrt{\c{6}(\bh - \c{24})}\right] \quad \mbox{\ for\ } \quad \frac{h \bh}{h + \bh} > \c{24} \ . 
\ee
This is identical to the result we derived from the free energy (\ref{CardyS}).

\end{spacing}

\end{document}